\documentclass[final]{svjour3}
\usepackage{graphicx}
\usepackage{rotating}
\usepackage{amssymb}
\makeatletter
\journalname{Journal of Low Temperature Physics}

\usepackage{comment}
\usepackage[labelfont=bf]{caption}
\usepackage{subcaption}
\usepackage{amsmath}
\usepackage[sort,numbers]{natbib}
\begin{document}

\newcommand{\hdblarrow}{H\makebox[0.9ex][l]{$\downdownarrows$}-}
\title{Parametric amplification via superconducting contacts in a Ka band niobium pillbox cavity}

\author{V. Gilles \and D. Banys \and M. A. McCulloch \and L. Piccirillo \and T. Sweetnam}

\institute{Jodrell Bank Centre for Astrophysics, The University of Manchester, \\  Manchester, M13 9PY,
United Kingdom\\ Tel.: +44 161 306 6470\\
\email{valerio.gilles@manchester.ac.uk}}

\maketitle

\begin{abstract}

Superconducting parametric amplifiers are commonly fabricated using planar transmission lines with a non-linear inductance provided by either Josephson junctions or the intrinsic kinetic inductance of the thin film.
\newline
However, Banys et al. \cite{Banys} reported non-linear behaviour in a niobium pillbox cavity, hypothesising that below $T_c$, the pair iris-bulk resonator would act as a superconducting contact surface exploiting a Josephson-like non-linearity. This work investigates this effect further by applying Keysight Technologies' Advanced Design System (ADS) to simulate the cavity using an equivalent circuit model that includes a user defined Josephson inductance component. The simulations show that for a resonance centred at $\nu_0=30.649$ GHz, when two tones (pump and signal) are injected into the cavity, mixing and parametric gain occur. The maximum achievable gain is explored when the resonator is taken to its bifurcation energy. These results are compared to cryogenic measurements where the pump and signal are provided by a Vector Network Analyzer.


\end{abstract}

\section{Introduction}
\begin{figure}[htbp]
\centering
\begin{subfigure}{.5\textwidth}
  \centering
\includegraphics[width=0.65\textwidth, keepaspectratio]{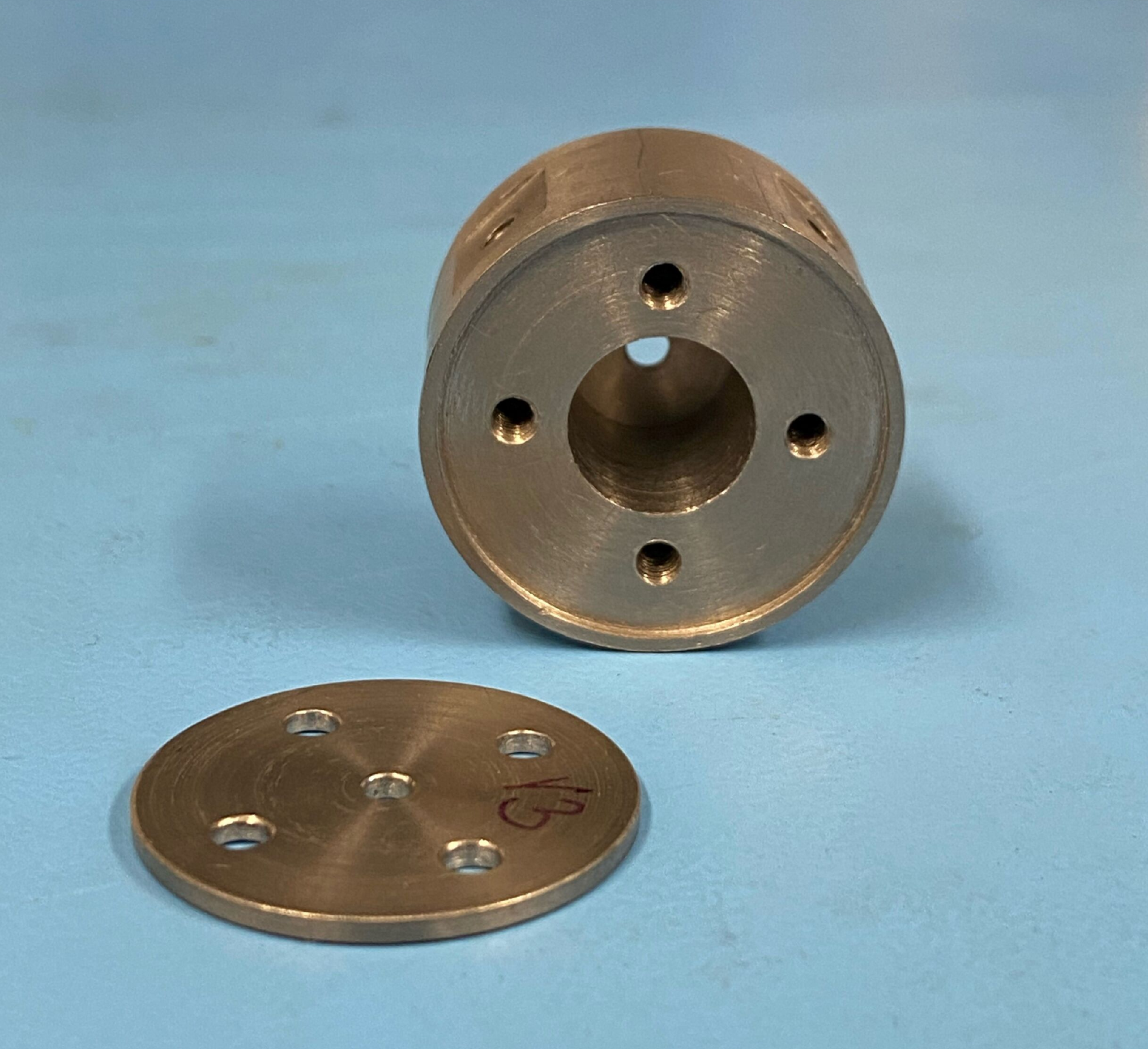}
  \caption{}
  \label{Figure1a}
\end{subfigure}%
\begin{subfigure}{.5\textwidth}
  \centering
  \includegraphics[width=1.0\textwidth, keepaspectratio]{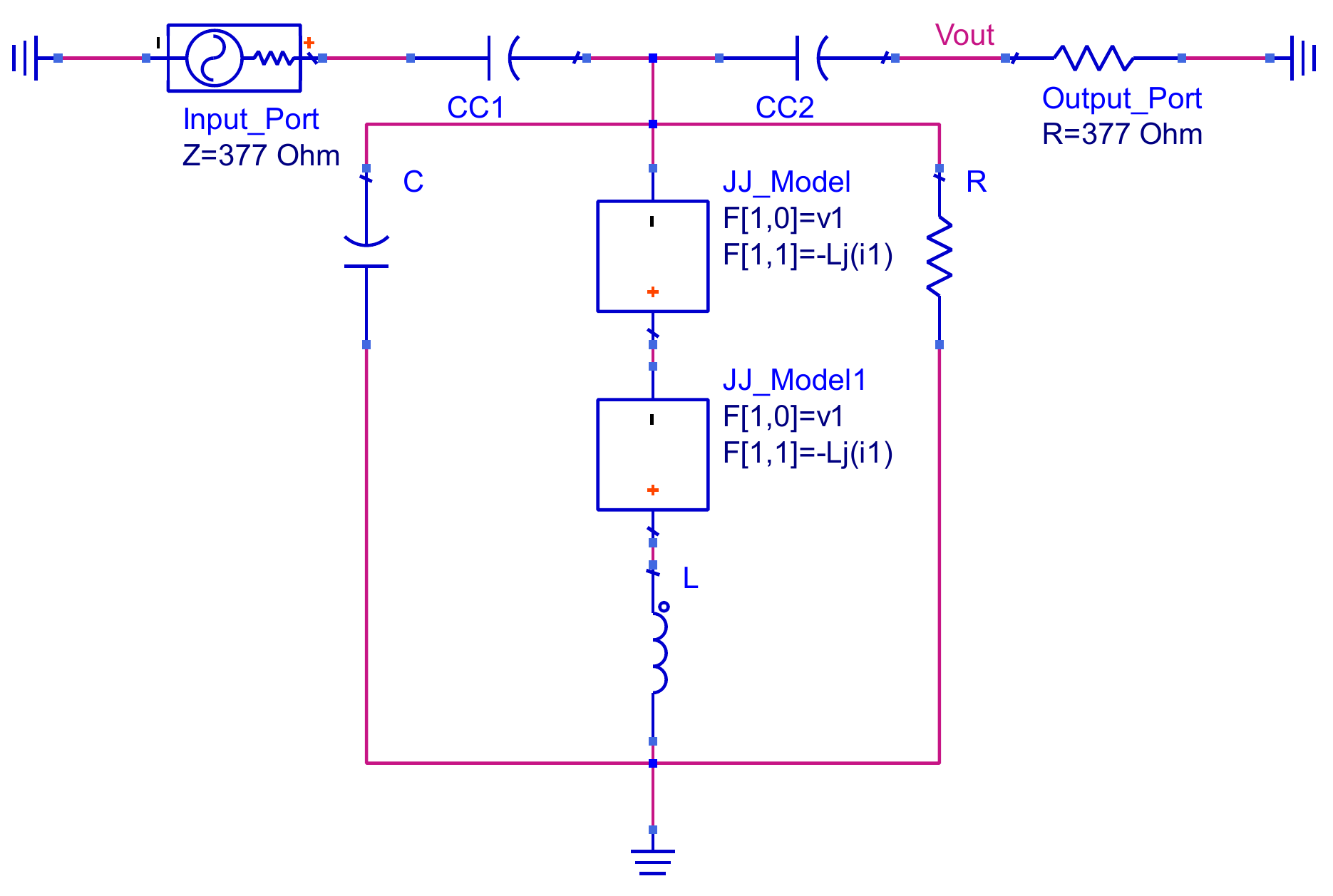}
  \caption{}
  \label{Figure1b}
\end{subfigure}
\caption{(a) - Picture of the two-ports cavity showing one of the irises. (b) - ADS schematic view of the niobium pillbox cavity as a parallel RLC circuit. The impedance of the input and output ports has been set to 377 $\mathrm{\Omega}$, approximately what is expected from the two lines of the Vector Network Analyzer (VNA). R, L and C are respectively the resistance, inductance and capacitance estimated for the TM$_{110}$ mode. CC1 and CC2 are two coupling capacitors and JJ$_{\textrm{Model}}$ and JJ$_{\textrm{Model1}}$ are the two symbolically defined devices containing the Josephson inductance, one for each iris}
\end{figure}

Interest in superconducting parametric amplifiers (SPAs) has grown rapidly over the last decade, due to their decent gain and the potential for quantum and sub-quantum noise performance \cite{PRXQuantum.2.010302}.
This noise characteristic gives them applications in a wide range of areas, from quantum computing  \cite{esposito2021perspective,9134828} and quantum communications \cite{doi:10.1126/science.1231930}, to research in fundamental physics and astrophysics \cite{DEJAGER1962266}.
The two most common types of SPAs are travelling-wave parametric amplifiers and resonator amplifiers (see \cite{esposito2021perspective} for a recent review).\\
To date the majority of SPAs have been designed to work below 20 GHz, but several works have shown that resonators can act as a good starting point for exploring parametric amplification at higher frequencies (see \cite{PhysRevApplied.13.024056} for a kinetic inductance example at $\sim$90 GHz). Another recent paper \cite{Banys} showed the possibility of achieving about 2 dB of parametric gain at $\sim$30 GHz, using a niobium pillbox cavity when brought near bifurcation. The authors observed signal gain when two tones (pump and signal) were fed into a TM$_{110}$ resonance and hypothesized that the main source of non-linearity could be the superconducting surface contacts formed by the irises and the bulk niobium structure (see Fig.~\ref{Figure1a}). Imperfections in these contacts would act as superconducting weak-links \cite{Abdo2006} which would provide a Josephson-like non-linearity.\newline

\noindent A Josephson junction (JJ) is a device made of two superconducting electrodes separated by either a non-superconductor (S-N-S junction), an insulator (S-I-S) or a weaker superconductor (S-s-S). A tunnelling current can flow across this interface, with a current-voltage relationship given by \cite{RevModPhys.46.251}:
\begin{equation}
\Delta V_j=\frac{L_j}{\textrm{cos}(\phi)}\frac{\partial I_j}{\partial t}
    \label{Equation1}
\end{equation}
where $\Delta V_j$ is the voltage difference across the electrodes, $I_j$ the current flowing through the junction,  $I_c$ the critical current and $L_j=\phi_0/(2\pi I_c)$ the zero-current Josephson inductance, which is analogous to the magnetic inductance. The current dependant Josephson inductance is defined as \cite{beltran2007}:
\begin{equation}
    L_j(I) = \frac{\hbar}{2eI_c}\frac{\arcsin{I_j/I_c}}{I_j/I_c}
    \label{Equation2}
\end{equation}
and, being non-linear with current, it can be used to facilitate mixing between two tones (pump and signal) producing gain, if the phase matching condition between these two tones is maintained. Josephson junctions have been widely used to make parametric amplifiers \cite{Zorin_2021,doi:10.1126/science.aaa8525,Eichler_2014} and often connected in series in order to increase the gain.

\section{Simulating the cavity in ADS}
\begin{figure}[htbp]
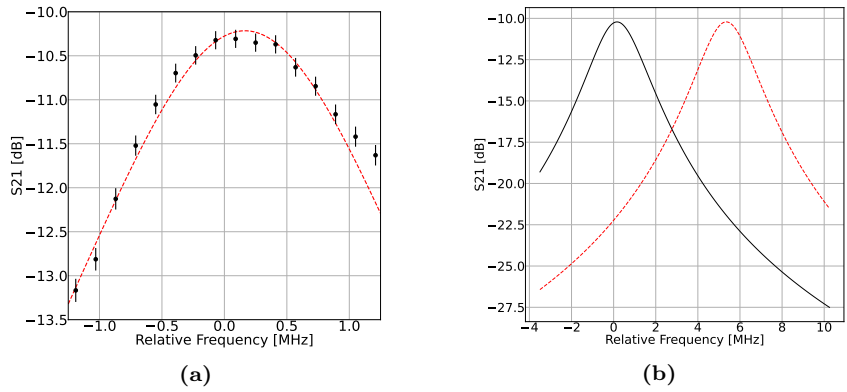

\centering
\begin{subfigure}{.5\textwidth}
  \centering
\includegraphics[width=0.8\textwidth, keepaspectratio]{Figure2a.pdf}
  \caption{}
  \label{Figure2a}
\end{subfigure}%
\begin{subfigure}{.5\textwidth}
  \centering
  \includegraphics[width=.8\textwidth, keepaspectratio]{Figure2b.pdf}
  \caption{}
  \label{Figure2b}
\end{subfigure}
\caption{(a) - (data points) Measured resonance centred at 30.649 GHz. (dotted line) - Same resonance obtained simulating the 2-ports cavity as a parallel RLC. (b) - Simulated resonance profiles centred at 30.649 GHz in the case of no input power (dotted line) and -60 dBm of input power (solid line)}
\end{figure}

This work further investigates mixing effects and parametric gain of a two-ports resonant niobium pillbox cavity made of annealed niobium (99.9$\%$ purity), around the resonance centred at $\nu_0\simeq30.649$ GHz. Two degenerate modes resonate at this frequency: the TM$_{110}$ (mode of interest) and the TE$_{010}$. The device is 15.7 mm long, has an inner diameter of 6 mm (see Fig.~\ref{Figure1a}) and an iris covering each port for improved transmission at Ka band and the decoupling of the two degenerate modes. The TM$_{110}$ mode is characterized by surface currents travelling alongside the length of the cavity, perpendicularly to the irises.\newline
The cavity has been simulated using ADS as a parallel RLC circuit (see Fig.~\ref{Figure1b}). ADS offers the possibility of using an harmonic balance solver which is key for this work and has been proven to be a very useful tool for simulating SPAs \cite{Sweetnam2022}. Other commonly used commercial software with already developed JJ models like WRSpice \cite{Delport} can simulate these circuits only via transient analysis (also due to limitations of the model), which is extremely inefficient for frequency sweeps. The symbolically defined device used (see Fig.~\ref{Figure1b}) contains two differential equations defined as: $\Delta V_j-\partial_t ^x \, F[1,x]=0$ ($x=0,1$). For $x=0$ the clear solution to the equation is: $F[1,0]=\Delta V_j$ while for $x=1$: $F[1,1]=-L_j(I)$, where $L_j=L_j(I)$ is the non-linear Josephson inductance (Eq. \ref{Equation2}).\\
Estimating the R, L and C values to use for the equivalent circuit model is not trivial. Initial values were calculated using the method described in \cite{Toman} and then fine tuned in order to get a resonance similar to the one measured for the TM$_{110}$ mode (Fig.~\ref{Figure2a}). The data points are $S_{21}$ measured at 1 K using a Agilent Technologies PNA-X Network Analyzer (10 MHz - 50 GHz) with an input power of -60 dBm, while the dotted line is the simulated resonance obtained with the tuned parameters.
The figure shows reasonably good matching between the two profiles for: $\mathrm{R}=6.36$ M$\mathrm{\Omega}$, $\mathrm{L}=2.80$ nH, $\mathrm{C}=9.19$ fF and $\mathrm{CC1}=\mathrm{CC2}=0.22$ fF (coupling capacitors).
Specifying the input power is important because of the excitation of the superconducting weak-links. In fact, as the input power is increased, the JJs are brought closer to their bifurcation energy, causing a shift in the resonance frequency. Unlike the zero-current Josephson inductance (Eq. \ref{Equation1}), the non-linear Josephson inductance (Eq. \ref{Equation2}) changes with current and the same happens to the frequency shift. Fig.~\ref{Figure2b} shows this effect: the dotted line corresponds to the case of negligible input power (zero non-linear Josephson inductance), while the solid line to -60 dBm of input power. There is a clear difference between the two cases, with a simulated frequency shift of about 6 MHz.\\
Another phenomenon observed with this cavity that can be power-dependant is the presence of "steps" or "discontinuities" in the resonance profile, similar to the ones reported in \cite{Abdo2006}. These weak-link features start to be relevant for an input power of -50 dBm and fade away above -25 dBm, causing abrupt changes to the transmission spectrum, while being very sensitive to magnetic fields.

\begin{figure}[htbp]
\centering
\includegraphics[width=0.85\textwidth]{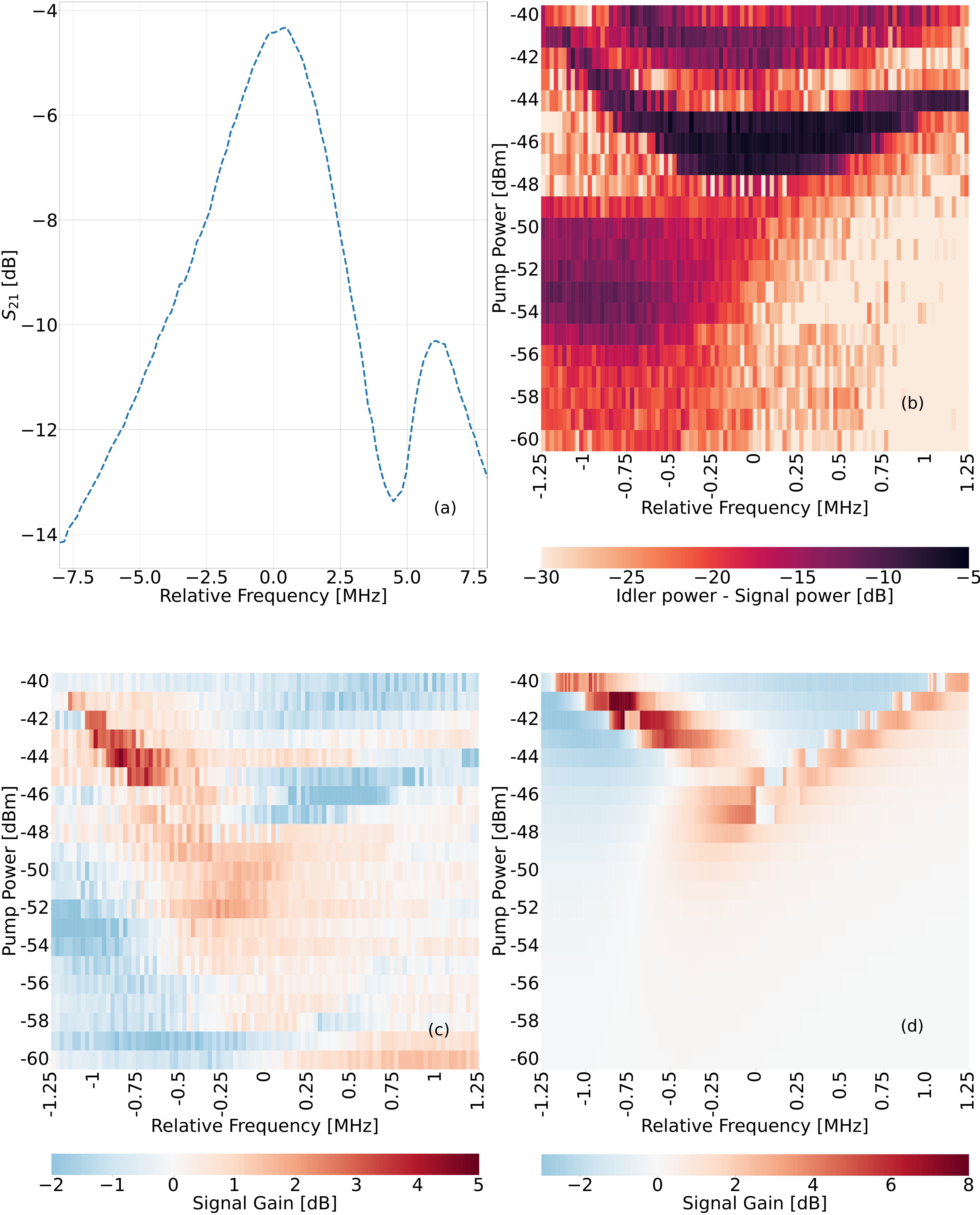}
\caption{(a) - $S_{21}$ measured at 1 K with an input power of -60 dBm, the relative frequency on the x axis is $f-30.643 $ GHz (centre of the TE$_{010}$ resonance). (b) Heat map showing $P_i-P_s$ (where $P_i$ is the power of the idler tone and $P_s$ the power of the signal tone). (c) - Heat map showing the measured signal gain as a function of the pump power and the relative frequency between the pump tone and the centre of the TM$_{110}$ resonance (on the x axis: $f_p-30.649$ GHz). The signal tone is kept 10 kHz behind the pump and at -60 dBm. (d) Same heat map showing $P_i-P_s$ (where $P_i$ is the power of the idler and $P_s$ the power of the signal). (d) Signal gain heat map simulated in ADS using a critical current of 1.5 mA}
\label{Figure3}
\end{figure}

\noindent To further investigate the findings reported in \cite{Banys} a $\mu$ - metal magnetic shield has been used and the gain measurements repeated at a lower temperature of $\sim$1 K by connecting the cavity to a single stage $^4$He sorption cooler. The quality factor of the TM$_{110}$ mode and the gain are expected to increase at lower temperatures due to the decrease in surface losses. Fig.~\ref{Figure3}(c) shows the measured signal gain as a function of the pump power (swept from -60 to -40 dBm), sweeping the pump frequency across a 2.5 MHz band centred at 30.649 GHz. The signal tone is kept 10 kHz behind the pump and at a power of -60 dBm. The figure shows a V-like gain pattern forming at the centre of the resonance and at a pump power of $\sim$ -50 dBm. The lack of magnetic shielding, the temperature difference and the different pressure applied to the irises when screwed on the cavity could explain the absence of this specific pattern during the previous studies \cite{Banys}. These patterns are hypothesised to be the bifurcation energies of the weak-links present between the superconducting contacts, attenuated by the transmission profile. A maximum signal gain of $\sim$ 5 dB has been measured from the cavity, about 3 dB more than what reported in \cite{Banys}.\\
Fig.~\ref{Figure3}(a) shows the transmission spectrum for an input power of -60 dBm (measured at 1 K) for a 16 MHz band centred at 30.643 GHz. The two resonances are believed to be the degenerate pair TE$_{010}$ (subject of the previous studies \cite{Banys}) and TM$_{110}$ split apart due to the decoupling provided by the irises. It is clear that the TM$_{110}$ mode (on the right) suffers lower transmission than the TE$_{010}$, probably due to the interaction between the two. For these studies, the interaction between modes has not been taken into account, therefore the transmission of the TM$_{110}$ mode has been boosted to match the other one. Of the two, the TM$_{110}$ is expected to show most of the gain, having surface currents flowing alongside the resonator, facilitating the tunnelling of Cooper pairs within the superconducting contacts. Other modes of propagation have been studied as well showing no significant mixing effects, proving that a current flow perpendicular to the surface contacts is necessary for signal gain. Tests using copper irises instead of niobium showed again no mixing effects, proving the concept that the superconducting pair iris-bulk resonator is indeed the main source of non-linearity.\\ 
Fig.~\ref{Figure3}(b) shows $P_i-Ps$ measured in analogous conditions and for an identical sweep of the parameters. A clear increase in the idler power is observed at about -47 dBm, where maximum mixing and gain are measured. Fig.~\ref{Figure3}(d) shows the same map as Fig.~\ref{Figure3}(c) but simulated in ADS using an equivalent circuit model (see Fig.~\ref{Figure1b}). The main non-linearity driving the gain is the current-dependant Josephson inductance included in the symbolically defined device, with the scaling parameter being the critical current (Eq. \ref{Equation2}). The bifurcation energy (and so the position of the V-patterns) should increase with the critical current while the overall gain should decrease due to a smaller Josephson inductance (Eq. \ref{Equation2}). Therefore, the simulations have been tuned in order to replicate the overall gain and position of the gain patterns seen in Fig.~\ref{Figure3}(c), obtaining a critical current of 1.5 mA. As it can be seen from Fig.~\ref{Figure3}(d), this value for the critical current replicates reasonably well the patterns measured from the cavity, while slightly overestimating the gain. The branch on the right in the simulated map seems to be more symmetrical than the measured one, likely due to the asymmetry (visible in Fig.~\ref{Figure2a}) of the measured resonance profile, that could not be reproduce without including the interaction between the two degenerate resonances.

\section{I-V curve measurements}
\begin{figure}[htbp]
\centering
\begin{subfigure}{.5\textwidth}
  \centering
\includegraphics[width=1\textwidth, keepaspectratio]{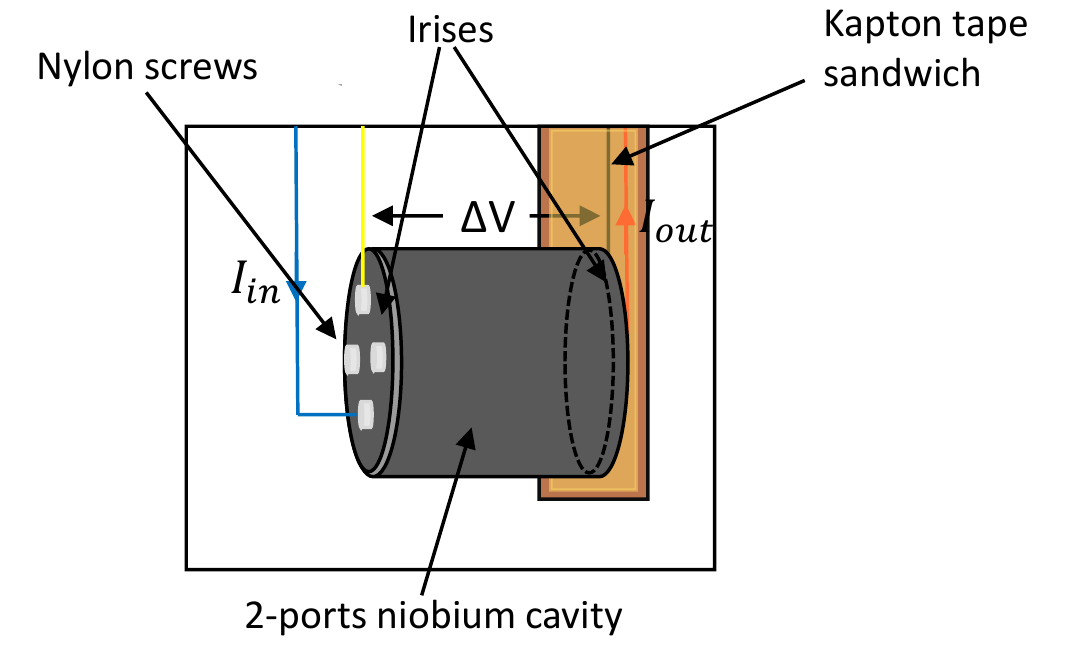}
  \caption{}
  \label{Figure4a}
\end{subfigure}%
\begin{subfigure}{.5\textwidth}
  \centering
  \includegraphics[width=.8\textwidth, keepaspectratio]{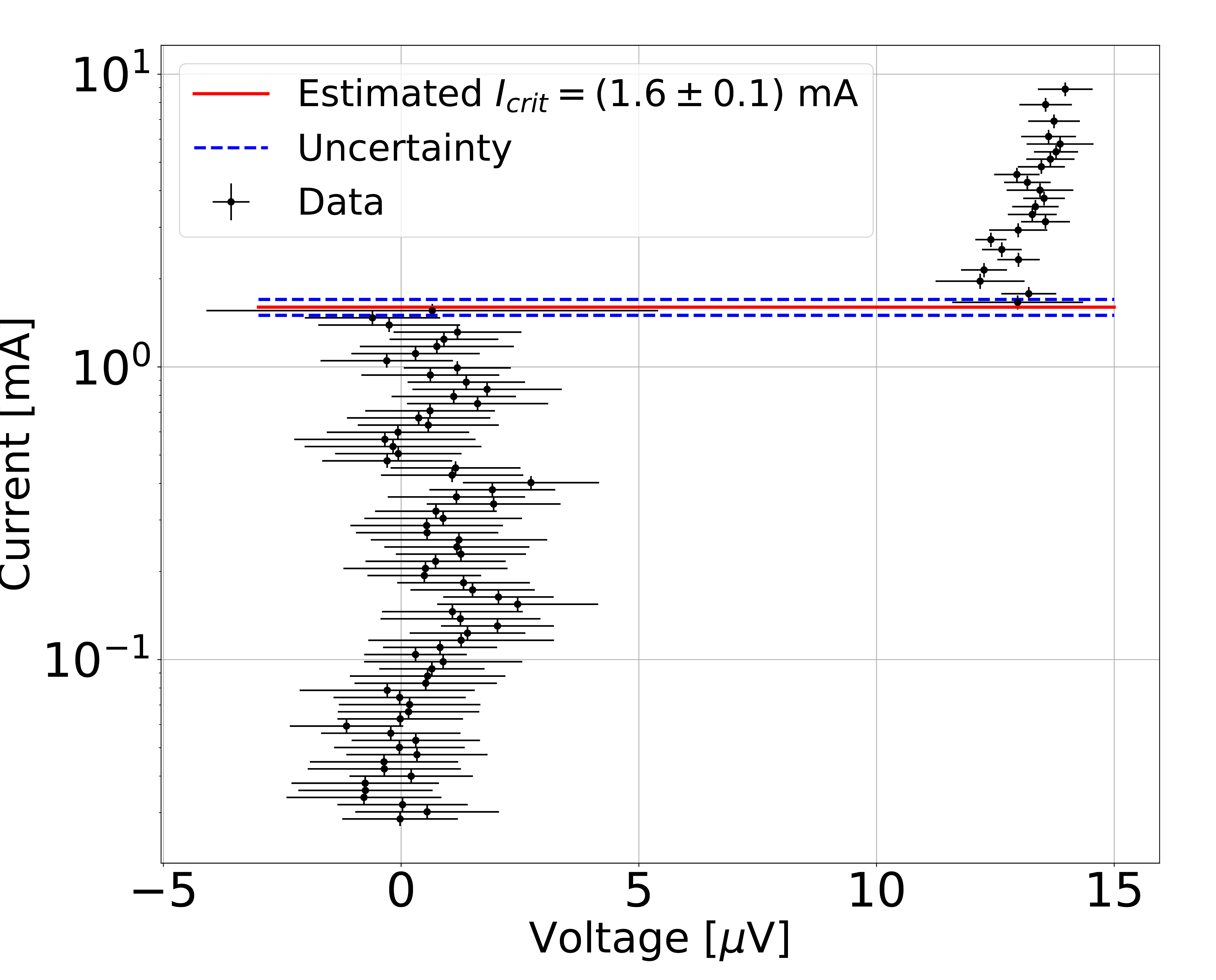}
  \caption{}
  \label{Figure4b}
\end{subfigure}
\caption{(a) - I-V curve measurement setup. Nylon screws have been used to attach the irises to the main body of the cavity in order to prevent the current from being lost through them. The cavity is attached to the 4 K plate of a cryostat cooled down by a Gifford-McMahon cryocooler. (b) - Logarithmic scale I-V curve of the 2-ports cavity measured at 6 K (the voltage measured is $\Delta V$ from Fig.~\ref{Figure4a}). The transition between the zero voltage Josephson regime and the Ohm regime is clearly visible}
\end{figure}

The previous section showed how a critical current of 1.5 mA is necessary in order to replicate the gain heat map patterns measured with the two-ports cavity.\\
If the superconducting surface contacts were actually acting as Josephson junctions, it should be possible to measure a Josephson-like I-V curve from it (for more information see \cite{Mitchell2012NiobiumDS,Shibata} and references therein). To measure this I-V curve, the setup shown in Fig.~\ref{Figure4a} has been adopted. To screen the wires from external radiation, twisted pairs surrounded by both a layer of aluminium and Kapton tape have been used. Outside the cryostat, coaxial and triaxial cables are used to lead the signal to the Keithley semiconductor characterization system (model 4200 - SCS). The measured I-V curve is shown in Fig.~\ref{Figure4b}. Each data point shown is an average of 20 points measured by the Keithley in a very small time interval, with the uncertainty assigned as $\sigma/\sqrt{20}$. The critical current $I_c=(1.6 \pm 0.1)$ mA has been estimated as $I_c=(I_{2}+I_{1})/2$ and its uncertainty as $\Delta I_c=(I_2-I_1)/2$ where $I_1$ and $I_2$ are respectively the current before and after the Josephson - Ohm transition. This value for the critical current agrees with the value found by tuning the simulations in Section 2. The temperature of the cavity ($\sim$6 K) has been measured using a Cernox sensor screwed on the main body. The thermal link used to connect it to the main plate (at $\sim 3$ K) was provided by a copper strap attached to one of the irises (Fig.~\ref{Figure4a}). Measurements performed at higher temperatures showed no change in critical current (while the gap voltage \cite{POOLE2014501} changed).
\vspace{1cm}

\section{Conclusions}
This work has presented tests and simulations conducted with the aim of explaining the non-linear effects measured with the two-ports niobium cavity. Using a parallel RLC circuit model including Josephson junctions in ADS, the performance of the device (measured at 1 K) has been simulated using a critical current: $I_c=1.5$ mA. This value agreed with the one estimated by measuring the Josephson I-V curve of the cavity: $I_c=(1.6\pm0.1)$ mA. These results proved that the parametric gain coming from the device can be attributed to the weak-links present between the superconducting contacts formed by the irises and the bulk niobium structure, exploiting a Josephson-like non-linearity.\\
These promising results suggest the possibility of developing a novel low-noise superconducting parametric amplifier based on series of superconducting contacts for increased non-linearity and gain.

\begin{acknowledgements}
"This project has received funding from the European Union’s Horizon 2020 research and innovation programme under the Marie Skłodowska-Curie grant agreement No 811312."
\\\\
\textit{The datasets generated during and/or analysed during the current study are available from the corresponding author on reasonable request.}
\end{acknowledgements}

\pagebreak


\bibliography{references}
\end{document}